\documentclass[aps,prd,showpacs,twocolumn]{revtex4}
\usepackage{amssymb}
\usepackage{amsmath}

\begin{document}

\title{Multipolar corrections for Lense-Thirring precession}

\author{Marcelo Zimbres}
\email{mzimbres@ifi.unicamp.br}
\affiliation{Instituto de F\'isica Gleb Wataghin, Universidade
 Estadual de Campinas,
Campinas  13083-970, S\~ao Paulo, Brazil}

\author{Patricio S. Letelier}
\email{letelier@ime.unicamp.br} 
\affiliation{Departamento de Matem\'atica Aplicada-IMECC, Universidade Estadual
de Campinas, Campinas  13083-970, S\~ao Paulo, Brazil}

\date{\today}

\begin{abstract} 
For stationary axially symmetric spacetimes we find a simple expression for the
Lense-Thirring precession in terms of the Ernst potential. This expression is
used to compute, in the weak field approximation, the major non-spherical
contributions to the  precession of a gyroscope orbiting the Earth. 
We reproduce  previously known results  and give a
new estimation for  non-spherical contributions.
\end{abstract}


\maketitle

\section{Introduction}

\label{sec:a}
 
The Gravity Probe B satellite launched in 2004 is a test of the general 
theory of relativity sensitive to off-diagonal components of the metric 
tensor resulting from the Earth rotation. The satellite contains 
a set of gyroscopes \cite{site} that, according to Einstein theory of gravity,
over the course of a year will precess about 6.606 arcsec/year(0.0018 degrees/year) 
in the orbital plane, an effect known as geodetic precession and 
39~milliarcsec/year(0.000011 degrees/year) in the plane of the Earth equator, 
an effect known as Lense-Thirring precession \cite{gravitation}\cite{ohanian}.
The experiment is now in its final phase of data analyze \cite{post}
and the results are intended to be published in the present year
\cite{site}. The Lense-Thirring part of the precession is the one related
directly to the off diagonal components of the metric tensor and so its
experimental verification will test the Einstein  theory for gravitation.

The aim of this paper is to relate  the expression for the Lense-Thirring precession 
with the mathematically sound theory of multipolar moments for axially symmetric
spacetimes due to Geroch~\cite{geroch} and Hansen~\cite{hansen}  and to
stimate the multipolar corrections based on some earth models. In the weak field
approximation we express our formulas in terms of Thorne moments \cite{thorne1} \cite{thorne2} 
using an harmonic coordinate system.  This is useful when we consider spacetimes whose
interior solutions are not known. The generalization to non linear fields
does not seem difficult and can be made using the full Thorne metric
\cite{gursel}.

The paper is outlined as follows, in Section~\ref{sec:b}, we
consider an axially symmetric stationary spacetime and use an appropriate tetrad
basis to obtain an exact expression for the Lense-Thirring precession. By using
the Thorne metric we specialize this  expression for the case of a weak
gravitational field. In Section~\ref{sec:c}  we discuss the precession  for the
Gravity Probe B experiment using three known  models for the Earth metric and a
model based in the exact solution of Einstein equation for a rotating mass with
a quadrupole moment. In particular, we discuss multipolar corrections. In
Section IV we summarize and discuss our results.

\section{The vector ${\bf \Omega}_{LT}$}
\label{sec:b}

Along the paper we use units such that  $c = G = 1$ and  metric signature  $(+,+,+,-)$. 
The symbol $<>$  denotes the symmetric and trace free part of a tensor.
$A_l$ is a shorthand notation for $a_{a_1}...a_{a_l}$. Greek letters
run from 1 to 4 and Latin from 1 to 3.

From the weak field approximation of  General Relativity  and the fact that the spin of an 
orbiting gyroscope is Fermi-Walker transported along its worldline \cite{synge}
we can show \cite{gravitation}\cite{ohanian} that the gyroscope spin will precess with an 
angular velocity ${\bf \Omega}$ that can be separated in three parts according 
to their physical origin,
\begin{align}
{\bf \Omega} = {\bf \Omega}_{T} + {\bf \Omega}_{DS} + {\bf \Omega}_{LT}, \label{om}
\end{align} 
where
\begin{align}
{\bf \Omega}_{T} = -\frac{1}{2}{\bf v}\times{\bf a}  \\ 
{\bf \Omega}_{DS} = \frac{3}{2}{\bf v}\times\nabla U  \\ 
{\bf \Omega}_{LT} = -\frac{1}{2}\nabla\times{\bf h}. 
\end{align}
The expression (\ref{om}) is general and it is  valid for any 
gyroscope describing a timelike world line.  The first term is known as the
Thomas precession, it depends on the gyroscope three-acceleration, ${\bf a}$,  as
well as, on the three-velocity, ${\bf v}$. It is null  when the gyroscope moves
along a geodesic. The second term is known as the geodetic or de Sitter precession
and is related to the Newtonian potential $U$. The third term  is the angular 
velocity associated to the Lense-Thirring effect. It is related to non-diagonal
part of the metric, ${\bf h} = g_{4j}{\bf e}_j$, where ${\bf e}_j$ denotes the 
spatial part of a tetrad base. An important characteristic of this term is that
it does not depend on the gyroscope velocity along the wordline. This fact
suggests to study the gyroscope  precession when it is at rest relative to a far
observer. Since $\bf v$  is null in such a frame the Thomas and the geodetic 
precession will vanish and we are left only with the Lense-Thirring precession.

The spacetime to be considered in this work  has the form,
\begin{multline}
ds^2 = g_{11}(dx^1)^2 + g_{22}(dx^2)^2 + g_{33}(dx^3)^2 + 2g_{34}dx^3dx^4  
\\ + g_{44}(dx^4)^2. \label{metrica}
\end{multline}
We will choose  the frame associated to the tetrads
\begin{eqnarray}
{\bf e}_{\hat{1}} &=& \frac{1}{\sqrt{g_{11}}}\frac{\partial}{\partial x^1} ,\ \
\ {\bf e}_{\hat{2}} = 
\frac{1}{\sqrt{g_{22}}}\frac{\partial}{\partial x^2}, \label{tetrad1} \\
{\bf e}_{\hat{3}} &= &\frac{1}{\sqrt{F}}\frac{\partial}{\partial x^3} -
\frac{g_{34}}{g_{44}}\frac{1}{\sqrt{F}}\frac{\partial}{\partial x^4}, \\
 {\bf e}_{\hat{4}}& =& \frac{1}{\sqrt{-g_{44}}}\frac{\partial}{\partial x^4},
\label{tetrad3}
\end{eqnarray}
where $F = g_{33} - g^2_{34}/g_{44}$ (see also \cite{mashhoon}). The
four velocity $u^{\mu}$ in this frame is given by $u^{\mu} = (0,0,0,(-g_{44})^{-1/2})$. 

To caculate the angular velocity of precession relative to the
choosen tetrads we use the Fermi-Walker transport law,
\begin{align}
{\bf \nabla}_{\bf u}{\bf S} = {\bf u}({\bf a}{\cdot} {\bf S}), \ \ \ {\bf a} \equiv {\bf \nabla}_{\bf u}{\bf u}.
\end{align}
Therefore
\begin{eqnarray}
\frac{dS_{\hat{j}}}{d\tau} &= &  {\bf \nabla}_{\bf u}({\bf S}{\cdot} {\bf e_{\hat{j}}}),\\
&=&{\bf S}{\cdot}({\bf \nabla}_{\bf u}{\bf e_{\hat{j}}}), 
\end{eqnarray}
where ${\bf \nabla}_{\bf u}$ is the total derivative operator, {\bf u} the
four-velocity and {\bf S} the spin vector. Using the fact that ${\bf
e}_{\hat\mu}{\cdot}{\bf e}_{\hat\nu} = \eta_{\hat\mu\hat\nu}$ and considering
that the spinning particle is at rest relative the tetrads we have ${\bf u}
= {\bf e}_4$. Hence the condition ${\bf{u}}\cdot{\bf S} = 0$ gives us
$S_{\hat{4}} = 0$.  From $\nabla {\bf e}_{\mu} = {\bf
e}_{\nu}{\bf m}^{\nu}_{\,\,\mu}$, we find that the Fermi-walker transport law
can be cast as,
\begin{eqnarray}
\frac{dS_{\hat{j}}}{d\tau} &=&({\bf S}\cdot{\bf e}_{\hat{\nu}})({\bf m}^{\hat{\nu}}_{\,\,\hat{j}}\cdot{\bf u}),\\ & =& 
S_{\hat{a}}({\bf m}^{\hat{a}}_{\,\,\hat{j}}\cdot{\bf e}_{\hat{4}}).
\end{eqnarray}
Now using the connection coefficients given in Appendix \ref{sec:e} and omitting
the subscript $LT$ in ${\bf \Omega}_{LT}$ we find,
\begin{align}
\frac{d{\bf S}}{d\tau} = {\bf \Omega}\times{\bf S}, 
\end{align}
where
\begin{align}
{\bf \Omega} = \Gamma_{\hat{2}\hat{3}\hat{4}} {\bf m}^{\hat{1}} + \Gamma_{\hat{3}\hat{1}\hat{4}}{\bf m}^{\hat{2}}.
\label{omega}
\end{align}
The symbols $\Gamma_{\hat{3}\hat{1}\hat{4}}$ and
$\Gamma_{\hat{3}\hat{2}\hat{4}}$ are given by
\begin{eqnarray}
\Gamma_{\hat{3}\hat{1}\hat{4}} & =&  \frac{g_{34}}{2\sqrt{-g_{44}g_{11}
F}}\left[\ln\left(\frac{g_{34}}{g_{44}}\right) \right]_{,1},\label{omega1} \\
\Gamma_{\hat{3}\hat{2}\hat{4}} & =& \frac{g_{34}}{2\sqrt{-g_{44}g_{22}
F}}\left[\ln\left(\frac{g_{34}}{g_{44}}\right) \right]_{,2} .\label{omega2}
\end{eqnarray}
We can write (\ref{omega1}) and (\ref{omega2}) in a simpler
way introducing the norm $\lambda$ and the twist $\omega_{\mu}$ of the time-like
Killing vector field $\xi^{\mu} = (0,0,0,1)$,
\begin{align}
&&\lambda = \xi_{\mu}\xi^{\mu} \label{norm} \\
&&\omega_{\mu} = \epsilon_{\mu\nu\alpha\beta}\xi^{\nu}\nabla^{\beta}\xi^{\alpha}.
\label{twist}
\end{align}
From  (\ref{metrica}),(\ref{norm}), (\ref{twist}) and 
(\ref{tetrad1})-(\ref{tetrad3}) we find that
$\Gamma_{\hat{2}\hat{3}\hat{4}}$ and $\Gamma_{\hat{3}\hat{1}\hat{4}}$ can be
written as,
\begin{align}
\Gamma_{\hat{2}\hat{3}\hat{4}} = \frac{1}{2\lambda}\omega_{\hat{1}}, \ \ \
\Gamma_{\hat{3}\hat{1}\hat{4}} = \frac{1}{2\lambda}\omega_{\hat{2}}.
\label{AAA}
\end{align}
From the vacuum Einstein equations we know  that $\omega_{\mu}
= \nabla_{\mu}\omega$ \cite{stephani}. Hence  the angular velocity $\Omega_{\hat{\mu}}$ can be
cast as,
\begin{align}
\Omega_{\hat{\mu}} = \frac{1}{2\lambda}\nabla_{\hat{\mu}}\omega
\label{angular_twist}.
\end{align}
The scalars $\lambda$ and $\omega$ are related to the Ernst potential, $\Gamma$, 
by the relation $\Gamma = -\lambda + i\omega$ \cite{ernst}. The expression
(\ref{angular_twist}) is particularly interesting  because it relates, in an exact manner,
the angular velocity of precession to the  two fundamental scalars of a stationary
spacetime. We were unable to find  this expression in the
literature.

To express the scalar $\omega$ in terms of Thorne moments using an
harmonic coordinate system we use the definition (\ref{twist})
and the Thorne metric for a weak gravitational field \cite{thorne2}\cite{thorne1}. This metric is given by,
\begin{eqnarray}
&& g_{\mu\nu} = h_{\mu\nu} +\eta_{\mu\nu,} \;\;\gamma_{\mu\nu} = h_{\mu\nu} -\frac{1}{2}\eta_{\mu\nu}\eta^{\alpha\beta}h_{\alpha\beta}, \\
&&\gamma_{44} = \frac{4M}{r} + \sum^{\infty}_{l = 2}(-1)^l\frac{4M_{A_l}}{l!}[r^{-1}]_{,A_l} ,\label{thorne-00}\\
&& \gamma_{4j} = -\frac{2\epsilon_{jpq}S_pn_q}{r^2} \nonumber \\
\nonumber \\  && \;\;\;\ -\sum^{\infty}_{l =
2}(-1)^l\frac{4l\epsilon_{jpq}S_{pA_{l - 1}}}{(l + 1)!}[r^{-1}]_{,qA_{l - 1}}
\label{thorne-0j} \\
&&\gamma_{jk} = 0, 
\end{eqnarray}
where $\eta_{\mu\nu}$ is the Minkowski metric. $M_{A_l}$ and $S_{A_l}$ are the mass and angular momentum multipole moments respectively. 
A useful expression  is,
\begin{align}
[r^{-1}]_{,A_l} = (-1)^l(2l - 1)!!\frac{N_{A_l}}{r^{l+1}}, \label{dev-r}
\end{align}
where $N_{A_l}= n_{a_1}...n_{a_l}$ and $n_a = x^a/r$.

Now from the previous equations  we find   that $\omega_a$
can be written as
\begin{align}
\omega_{a} = \epsilon_{abc}\gamma_{4c,b}.
\end{align}
From Eqs. (\ref{thorne-0j}) and (\ref{dev-r}) get
\begin{align}
\omega = \sum^{\infty}_{l = 1}\frac{4l(2l-1)!!}{(l +
1)!}\frac{S_{A_l}N_{A_l}}{r^{l+1}}. 
\end{align}
 Now we shall  write this
solution in spherical coordinates. Since we are working in an axially symmetric
spacetime  our multipole moments $M_{A_l}$ and $S_{A_l}$ are multiples of
$\hat{z}^{<A_l>}$, the symmetric, trace-free outer product of the axis-vector
with itself.  The moments   are  completely determined by the numbers $M_l$ and
$S_l$ \cite{hansen} defined by
\begin{align}
M_l = M_{A_l}\hat{z}^{A_l}, \ \ \ S_l = S_{A_l}\hat{z}^{A_l}.
\end{align}
We can show that,
\begin{align}
M_{A_l} = \frac{(2l - 1)!!}{l!}M_l\hat{z}^{<A_l>}, \ \ \ S_{A_l} = \frac{(2l - 1)!!}{l!}S_l\hat{z}^{<A_l>}. 
\end{align}
To compute  the quantity $S_{A_{l}}N_{A_{l}}$ we use $S_{A_{l}}N_{A_{l}} = S_lP_l(\cos\theta)$, where $P_l(\cos\theta)$ are the usual Legendre polynomials. Hence,
\begin{align}
\omega = \sum^{\infty}_{l = 1}\frac{4l(2l - 1)!!}{(l +
1)!}\frac{S_lP_l(\cos\theta)}{r^{l+1}}.
\end{align}
To be consistent in our approximation we use  $g_{44} = -1$ in (\ref{angular_twist}).  Then
\begin{align}
\Omega_{\mu} = -\frac{1}{2}\nabla_{\mu}\omega. \label{angular_twist2}
\end{align}
Therefore in spherical coordinates,
\begin{multline}
{\bf \Omega} = \sum^{\infty}_{l = 1}\frac{2l(2l - 1)!!}{(l + 1)!}\frac{S_l}{r^{l+2}}[(l + 1)P_l(\cos\theta)\hat{r} \\
 + \sin\theta P^{'}_l(\cos\theta)\hat{\theta}], \label{tey}
\end{multline}
where $()'$ denotes derivative with respect to $\cos\theta$.
Equation (\ref{tey}) shows that, in this approximation, the contributions to ${\bf \Omega}$ are
only from the angular momentum moments. In  Appendix~\ref{sec:g} we show that
the above formula is equivalent to formula (23) of 
\cite{teyssandier1}. Relating  ${\bf \Omega}$  to the Cartesian unit vectors 
$\hat{x}$, $\hat{y}$ and $\hat{z}$, we get
\begin{align}
{\bf \Omega} = \sum^{\infty}_{l = 1}\frac{2l(2l - 1)!!}{(l + 1)!}\frac{S_l}{r^{l+2}}{\bf f}(l,\theta), \label{omega-cartesiano}
\end{align}
where the components of ${\bf f}(l,\theta)$ are
\begin{eqnarray}
f^x& =& (l + 1)P_l(\cos\theta)\sin\theta 
+ P^{'}_l(\cos\theta)\cos\theta\sin\theta, \\
f^y& =& 0,\\
f^z &=& (l + 1)P_l(\cos\theta)\cos\theta - P^{'}_l(\cos\theta)\sin^2\theta .
\end{eqnarray}
In deriving the previous expressions, for convenience, it was  assumed that the orbit is on the plane $y = 0$.

\section{ {$\bf \Omega$} for the gyroscope experiment}

\label{sec:c}

The ideal orbit for the Gravity Probe B satellite is a  circular 
one with altitude $642$km. Since the  
Earth has non-vanishing multipole moments this circle is slightly distorted. 
 The  radius of such  orbit in the lowest order correction of the quadrupole moment is given in 
\cite{barker2}. In terms of Thorne  quadrupole moment $Q$ the orbit can be written as,
\begin{align}
r = r_0\left(1 - \frac{3Q}{8M}\frac{\cos2\theta}{r^2_0}\right),
\end{align}
where $M$ and $Q$ are the mass and Thorne quadrupole moment of the source, respectively. From  Eq. (\ref{omega-cartesiano}),  expanding in powers of $1/r_0$,  we obtain
\begin{align}
{\bf \Omega} = \frac{S}{2r_0^3}\left[{\bf g}_1(\theta) - \frac{27Q}{16Mr_0^2}\left({\bf g}_2(\theta) - \frac{5MS_3}{2QS}{\bf g}_3(\theta)\right)\right], \label{ave}
\end{align}
where the vectors ${\bf g}_j(\theta),\ \  j=1,2,3$ are,
\begin{align}
{\bf g}_1 = 2{\bf f}(1,\theta) \ \ \ {\bf g}_2 = \frac{4}{3}{\bf f}(1,\theta)\cos2\theta, \ \ \ {\bf g}_3 = \frac{16}{9}{\bf f}(3,\theta).
\end{align}
The vectors ${\bf g}_j$ have the property $\langle g^x_j\rangle = \langle g^y_j\rangle = 0$ and $\langle g^z_j\rangle = 1,$ where $\langle g^i_j\rangle$ means to take the average over $g^i_j(\theta)$, which in this case is a simple integration over $\theta$ and in view of the orbit symmetry, it is enough to consider  half orbit from one pole to the other  $[0,\pi]$. 
The average of ${{\bf \Omega}}$ reads,
\begin{align}
\langle\overrightarrow{\Omega}\rangle = A\left[1 - B\left(1 -
\frac{5C}{2}\right)\right]\hat{z}, \label{omega-final}
\end{align}
where
\begin{align}
A = \frac{S}{2r^3_0}, \ \ \ B = \frac{27Q}{16Mr^2_0}, \ \ \ C = \frac{MS_3}{QS}.
\nonumber
\end{align}
The case $C = 2/5$ is special, since we have no correction although
the spacetime is deformed.

The Thorne mass moments can be accurately computed
using de data published in~\cite{geodetic}, but the angular momentum moments are
independent of the mass moments. To determinate the constant $C$ we need a
model for an oblate Earth. Some authors~\cite{adler}~\cite{teyssandier1} have
already calculated the non-spherical contributions to ${\bf \Omega}$. We shall use our 
approach to compare the precession for  these  models. Also, based on the 
Earth model of reference~\cite{adler2},  we give a new estimative for ${\bf \Omega}$.
\subsection{Teyssandier model}

Comparing the general form of Thorne metric with the one  in
\cite{teyssandier1}, we get
\begin{align}
Q = -\frac{2MR^2J_2}{3}, \ \ \ S_3 = -\frac{4SR^2K_2}{5}, \ \ \  C = 0.97.
\end{align}
where we have used the values $J_2 = (1082.64 \pm 0.01)\times10^{-6}$ e $K_2 =
0.874\times10^{-3}$ given by the author. Therefore  we will have
\begin{align}
\langle\overrightarrow{\Omega}\rangle = A(1 + 1.42B)\hat{z}. \label{teyssandier}
\end{align}
This value is different from the one in \cite{teyssandier1}
because we use the angle $\theta$ in~(\ref{omega-cartesiano}) instead of~$\psi
= \pi/2 - \theta$ as in the quoted reference.  Therefore on averaging
over~(\ref{ave}) we get a different, but equivalent result.

\subsection{ Adler-Silbergeit model}

For Adler-Silbergeit model B~\cite{adler}, we obtain
\begin{align}
Q = -\frac{2MR^2J_2}{3}, \ \ \ S_3 = -\frac{8\omega MR^4J_2}{35}, \ \ \ S =
I\omega.
\end{align}
From this expression  we calculate the constant $C = MS_3/QS$, and Eqs~(\ref{omega-final}) 
gives us,
\begin{align}
\langle\overrightarrow{\Omega}\rangle = A\left[1 - B\left(1 -
\frac{6}{7}\frac{MR^2}{I}\right)\right]\hat{z}.
\end{align}
Confronting this result with Eq.~(61) of \cite{adler}  we see that our third term differs
by a factor two (the numerical factor in the quoted reference  is 3/7). Using
$MR^2/I = 3.024$ given by the authors we obtain
\begin{align}
\langle\overrightarrow{\Omega}\rangle = A(1 + 1.59B)\hat{z}.
\end{align}
They  get $\langle\overrightarrow{\Omega}\rangle = A(1 + 0.30B)\hat{z}$.

\subsection{ Adler model}

For this model~\cite{adler2} we have
\begin{align}
Q = -\frac{2Ma^2}{9}, \ \ \ S_3 = -\frac{4Sa^2}{25}, \ \ \  C = 0.72.
\end{align}
Hence  the averaged angular velocity  is
\begin{align}
\langle\overrightarrow{\Omega}\rangle = A(1 + 0.80B)\hat{z}. \label{nova}
\end{align}

\subsection{ Exact solution}

It is instructive to calculate the constant $C$ for the case our geometry
is described by an exact solution of the  Einstein vacuum equations with
arbitrary quadrupole moment.  We choose the version of this solution
presented in~\cite{manko}. Equivalent  solutions has been obtained by several
authors using different methods~\cite{stephani}. The first four nonzero
Geroch-Hansen moments are,
\begin{eqnarray}
M_0 & = & k(1+\alpha^2)/(1-\alpha^2) \nonumber \\
J_1 & = & -2\alpha k^2(1+\alpha^2)/(1-\alpha^2)^2 \nonumber \\
M_2 & = & -k^3[\beta + 4\alpha^2(1+\alpha^2)(1-\alpha^2)^{-3}] \\
J_3 & = & 4\alpha k^4 [\beta
+ 2\alpha^2(1+\alpha^2)(1-\alpha^2)^{-3}]/(1-\alpha^2) \nonumber
\end{eqnarray}
where $M_n, J_n$ are  the mass and current moments, respectively,  and $\alpha,\beta$ and $k$ are
parameters. We find the following relation,
\begin{eqnarray}
J_3 = \frac{J}{M}(2M_2 - M_2^{Kerr})
\end{eqnarray}
where $M_2^{Kerr} = -J^2/M$ is the Kerr quadrupole moment. Since $M_2^{Kerr}$
has vanishing Newtonian limit, in our approximation, we are only left with
\begin{eqnarray}
J_3 = \frac{2M_2J}{M}.
\end{eqnarray} 
Now using the correspondence between Geroch-Hansen and Thorne moments~\cite{gursel}
we get  $C = 4/15$. Hence  equation~(\ref{omega-final}) reads
\begin{align}
\langle\overrightarrow{\Omega}\rangle = A(1 - 0.33B)\hat{z}.
\end{align}
that differs considerably from the previous models.

\section{Summary and Conclusions}
By using an adequate tetrad basis we derive an exact expression for the
Lense-Thirring precession in terms of the norm and the twist of the spacetime 
timelike killing vector. We  calculated the imaginary part of Ernst potential
using the weak field approximation and the Lense-Thirring precession
to  any desired multipolar correction. We consider the case of the Gravity
Probe B experiment expanding the precession  angular velocity  up to the  order
of $1/r_0^5$ and averaging over a half trajectory. The final  expression is
given in terms of Thorne multipole moments. In our opinion, this general 
expression, may be  interesting  for the  interpretation of exact solutions of 
Einstein field Equations. We have reproduced  known results and given new
estimatives for the non-spherical contributions to  precession  angular
velocity. We conclude that the earth model plays a significant role in the
multipolar contribution. The results produced by different Earth models are
accounted for by our constant C, fact that makes our relation
(\ref{omega-final}) particularly usefull.

\acknowledgments
The work of MZ was supported by Capes. 
PSL  thanks CNPQ and FAPESP for partial financial support.

\appendix

\section{Connection coefficients}

\label{sec:e}
For easy reference we give in this appendix the connection coefficients
relative to the tetrads (\ref{tetrad1})-(\ref{tetrad3}). The dual base is:
\begin{eqnarray}
{\bf m}^{\hat{1}}&=&\sqrt{g_{11}}dx^1, \  {\bf{m}}^{\hat{2}} =
\sqrt{g_{22}}dx^2, \\
{\bf{m}}^{\hat{3}}& =& \sqrt{F}dx^3 ,\  {\bf{m}}^{\hat{4}} =
\frac{g_{34}}{\sqrt{-g_{44}}}dx^3 - \sqrt{-g_{44}}dx^4.
 \end{eqnarray}
From the first Cartan structure equations~\cite{stephani},
\begin{align}
{\bf d}{\bf{m}}^{\mu} + {\bf{m}}^{\mu}_{\,\,\nu}\times{\bf{m}}^{\nu} =0,
\end{align}
we find the connection coefficients,
\begin{eqnarray}
{\bf{m}}_{\hat{1}\hat{2}}&=&\Gamma_{\hat{1}\hat{2}\hat{1}}{\bf{m}}^{\hat{1}} +  \Gamma_{\hat{1}\hat{2}\hat{2}}{\bf{m}}^{\hat{2}},\\
{\bf{m}}_{\hat{1}\hat{3}} &=& \Gamma_{\hat{1}\hat{3}\hat{3}}{\bf{m}}^{\hat{3}} +   \Gamma_{\hat{1}\hat{3}\hat{4}}{\bf{m}}^{\hat{4}} ,\\ 
{\bf{m}}_{\hat{2}\hat{3}} &=& \Gamma_{\hat{2}\hat{3}\hat{3}}{\bf{m}}^{\hat{3}} +  \Gamma_{\hat{2}\hat{3}\hat{4}}{\bf{m}}^{\hat{4}},\\
{\bf{m}}_{\hat{4}\hat{1}} &=& \Gamma_{\hat{4}\hat{1}\hat{3}}{\bf{m}}^{\hat{3}} +   \Gamma_{\hat{4}\hat{1}\hat{4}}{\bf{m}}^{\hat{4}}, \\ 
{\bf{m}}_{\hat{4}\hat{2}}& =& \Gamma_{\hat{4}\hat{2}\hat{3}}{\bf{m}}^{\hat{3}} +  \Gamma_{\hat{4}\hat{2}\hat{4}}{\bf{m}}^{\hat{4}},\\
{\bf{m}}_{\hat{4}\hat{3}}& = &\Gamma_{\hat{4}\hat{3}\hat{1}}{\bf{m}}^{\hat{3}} +   \Gamma_{\hat{4}\hat{3}\hat{2}}{\bf{m}}^{\hat{2}},
\end{eqnarray}
where
\begin{eqnarray}
\Gamma_{\hat{1}\hat{2}\hat{1}}&=& \frac{g_{11,2}}{2g_{11}\sqrt{g_{22}}},\ \ \Gamma_{\hat{2}\hat{1}\hat{2}} = \frac{g_{22,1}}{2g_{22}\sqrt{g_{11}}},\\
\Gamma_{\hat{3}\hat{1}\hat{3}} &= &\frac{F_{,1}}{2F\sqrt{g_{11}}}, \ \ \Gamma_{\hat{3}\hat{2}\hat{3}} = \frac{F_{,2}}{2F\sqrt{g_{22}}},\\
\Gamma_{\hat{1}\hat{4}\hat{4}} &=& \frac{g_{44,1}}{2g_{44}\sqrt{g_{11}}},\ \
\Gamma_{\hat{4}\hat{2}\hat{4}} = \frac{g_{44,2}}{2g_{44}\sqrt{g_{22}}} ,\\
\Gamma_{\hat{3}\hat{1}\hat{4}}& =&  \frac{g_{34}}{2\sqrt{-g_{44}g_{11} F}}\left[\ln\left(\frac{g_{34}}{g_{44}}\right) \right]_{,1} \\
\Gamma_{\hat{3}\hat{2}\hat{4}}& =& \frac{g_{34}}{2\sqrt{-g_{44}g_{22} F}}\left[\ln\left(\frac{g_{34}}{g_{44}}\right) \right]_{,2} 
\end{eqnarray}
and 
\begin{eqnarray}
\Gamma_{\hat{4}\hat{2}\hat{3}} = \Gamma_{\hat{3}\hat{2}\hat{4}}, \ \  \Gamma_{\hat{4}\hat{3}\hat{1}} =
\Gamma_{\hat{1}\hat{3}\hat{4}}, \\
 \Gamma_{\hat{4}\hat{1}\hat{3}} = \Gamma_{\hat{3}\hat{1}\hat{4}}, \ \ 
\Gamma_{\hat{4}\hat{3}\hat{2}} = \Gamma_{\hat{2}\hat{3}\hat{4}}. 
\end{eqnarray}

\section{Equivalent expressions}

\label{sec:g}

In this appendix we show that our Eq.~(\ref{tey}) is equivalent to Eq . (23) of 
Ref.  \cite{teyssandier1}. Using the relation,
\begin{align}
\hat{\theta} = \frac{1}{\sin\theta}(\cos\theta\hat{r} - \hat{z}), 
\end{align}
we can write equation~(\ref{tey}) in terms of the vectors~$\hat{r}$~and~$\hat{z}$, 
\begin{multline}
{\bf \Omega} = \sum^{\infty}_{l = 1}\frac{2l(2l - 1)!!}{(l + 1)!}\frac{S_l}{r^{l+2}}[(l + 1)P_l(\cos\theta)\hat{r} 
\\ + \cos\theta P^{'}_l(\cos\theta)\hat{r} - P^{'}_l(\cos\theta)\hat{z}].\label{01}
\end{multline}
From the relation, 
\begin{align}
P^{'}_{l + 1}(\cos\theta) = (l + 1)P_l(\cos\theta) + \cos\theta P^{'}_{l}(\cos\theta),
\end{align}
we can cast~(\ref{01}) in the form,
\begin{align}
{\bf \Omega} = \sum^{\infty}_{l = 1}\frac{2l(2l - 1)!!}{(l + 1)!}\frac{S_l}{r^{l+2}}
[P^{'}_{l + 1}(\cos\theta)\hat{r} - P^{'}_{l}(\cos\theta)\hat{z}],
\end{align}
or as,
\begin{multline}
{\bf \Omega} = \frac{S}{r^3}(3\cos\theta\hat{r} - \hat{z} + \\
\sum^{\infty}_{l = 1}\frac{2l(2l + 1)!!}{(l + 2)l!}\frac{(S_l/S)}{r^{l}}[P^{'}_{l + 1}(\cos\theta)\hat{r} - P^{'}_{l}(\cos\theta)\hat{z}]).
\end{multline}
Identifying,
\begin{align}
S_{l+1} = -\frac{(l+2)!SK_lR^l}{2(2l + 1)!!},
\end{align}
we get Eq.~(23) of Teyssandier paper.

\

\end{document}